\documentclass[showpacs,aps,prd,nofootinbib,floatfix,amsmath,amssymb]{revtex4}
\usepackage{graphicx}
\begin{document}


\title{Implications of a very light pseudoscalar boson on lepton flavor
violation}
\author{Adriana Cordero-Cid}
\author{G. Tavares-Velasco}
\author{J. J. Toscano}\affiliation{Facultad de
Ciencias F\'\i sico Matem\'aticas, Benem\'erita Universidad
Aut\'onoma de Puebla, Apartado Postal 1152, Puebla, Pue., M\'exico}

\date{\today}

\begin{abstract}
A long-lived very light pseudoscalar boson would favor lepton flavor
violating transitions of charged leptons. Its implications on the
$l_i\to l_j\gamma\gamma$, and $l_i\to l_je^+e^-$ transitions are
investigated. Assuming that $2m_e<m_\phi<m_\mu$, it is found that
the inequality $B(l_i\to l_j\gamma \gamma)<B(l_i\to l_je^+e^-)$ is
hold. The experimental constraints on the decays $l_i \to
l_j\gamma$, $l_i\to l_j l_k l_k$, and $l_i \to l_j\gamma\gamma$ are
used to bound the $\phi l_il_j$ couplings.
\end{abstract}
 \maketitle

In the standard model (SM) the couplings of the Higgs boson to the
remaining massive particles are thoroughly determined, which should
be considered an outstanding feature of the model. This allowed the
CERN large electron positron (LEP) collider to conclude its
operative stage with a significantly strong lower bound on the Higgs
boson mass of around $115$ GeV \cite{HBLEP}. However, as long as
additional scalars are included in the theory, the Higgs boson
masses become more difficult to bound due to the proliferation of
free parameters. The bonus is the appearance of interesting new
physics effects such as lepton flavor violating (LFV) and flavor
changing neutral currents (FCNC) transitions, which can be mediated
by the new Higgs bosons at the tree level. In particular, there are
well motivated theoretical arguments that favor the existence of a
very light scalar or pseudoscalar particle, which would have
remained undetected so far because it would interact very weakly
with ordinary matter. Since the Higgs boson masses are typically of
the same order of the Fermi scale, extended scalar sectors do not
lead automatically to the presence of light scalars, unless an
unnatural fine tuning is implemented between all the parameters of
the Higgs potential. However, an exception occurs if the theory
possesses an approximate global symmetry. If such a symmetry exits
and is spontaneously broken, a massless Goldstone boson arises,  but
if it is only approximate, a massive state arises, which is
naturally light. This symmetry is only approximate because a small
explicit symmetry breaking can be introduced in the classical
Lagrangian or generated through quantum effects such as anomalies.
The most common example is the axion \cite{AXIONS}, which is
associated with the spontaneously broken Peccei-Quinn symmetry
\cite{PQ}. Other known examples are familons \cite{FAM} and Majorons
\cite{MAJ}, which are associated with spontaneously broken family
and lepton number symmetries, respectively. One example of a global
symmetry broken explicitly through a small parameter in the Higgs
potential is the two-Higgs doublet model (THDM) \cite{SHER}, where
the pseudoscalar boson $A$ acquires a mass proportional to the small
parameter: $m^2_A=-\lambda_5 v^2$, with $v=246$ GeV the Fermi scale
\cite{SHER}. In this case, the Higgs potential possesses an exact
global $U(1)\times U(1)$ symmetry in the limit of vanishing
$\lambda_5$. As pointed out in Refs. \cite{TAVARES,SHER}, this
particle has a very interesting phenomenology if its mass is below
$200$ MeV.

We are interested in studying some phenomenological implications of
LFV transitions mediated by a very light pseudoscalar boson $\phi$,
with $2m_e<m_\phi<m_\mu$. We will focus on the three-body
transitions $l_i\to l_j\gamma \gamma$ and $l_i\to l_j e^+e^-$, with
$l_i=\mu, \tau$ and $l_j=e,\mu$. Although Higgs-mediated LFV effects
have long attracted considerable attention \cite{SHER1,SHER2}, the
current evidences of nonzero mass for the neutrinos \cite{FUKUDA}
has renewed the interest in this issue, and particularly in the
potential role that spin zero particles may play
\cite{RECENTWORKS,SHALOM}. In our case, the key ingredient is the
$\phi$ mass range, which imposes severe kinematical restrictions on
the possible $\phi$ channel decays. In fact, in this mass range, the
only kinematically-allowed tree-level decay mode is $\phi\to
e^+e^-$, though the one-loop induced mode $\phi \to \gamma \gamma$
can also be competitive since a highly suppressed $\phi e^+e^-$
vertex is expected in general. Notice that, for such a light $\phi$,
the LFV decays  $\phi \to l_il_j$ are not kinematically allowed.
Although the relative significance of each channel would be
model-dependent generally, it is also clear that the total decay
width $\Gamma_\phi$ would be very small. This means that such a
$\phi$ scalar boson is a long-lived particle. Our main motivation is
thus that a particle with such peculiarities can naturally favor LFV
transitions, not only because its mass is much smaller than any
heavy lepton, but also because it is a very long-lived particle. A
very long-lived $\phi$ has the property that $\Gamma_\phi \ll
m_\phi$, so the narrow-width approximation can be used when
calculating the three-body decays $l_i\to l_j\gamma \gamma$ and
$l_i\to l_j e^+e^-$. These facts, together with the fact that $\phi
\to e^+e^-$ and $\phi\to \gamma \gamma$ are the only decay modes of
$\phi$, will allow us to establish an interesting relation between
the $l_i\to l_j\gamma \gamma$ and $l_i\to l_j e^+e^-$ transitions.

Once the main motivations of this work were discussed, we proceed to
calculate the decay rates for the above mentioned channels. It is
well known that a Yukawa sector associated with an extended Higgs
sector predicts somewhat general couplings of the Higgs bosons to
the fermions. One of the most interesting features of such a scalar
sector is the presence of nondiagonal interactions both in the
lepton sector and the quark sector. In the context of a general
renormalizable theory, we assume that the couplings of the
pseudoscalar $\phi$ to the leptons are naturally suppressed by
introducing the Cheng-Sher ansazt \cite{SHER1}:
\begin{equation}
\lambda_{ij}\frac{\sqrt{m_im_j}}{v}\gamma_5,
\end{equation}
where $\lambda$ is a dimensionless nondiagonal matrix defined in the
flavor space. For scalar fields with masses of the order of the
Fermi scale, it seems reasonable to assume that $\lambda_{ij}\sim
O(1)$, but this may be unnatural for a very light $\phi$ boson. As
will be seen below, this is the case for transitions involving the
two first lepton families since the current experimental constraints
on $\mu \to e\gamma$  indicate that $|\lambda_{\mu e}|\ll 1$. Below
we will consider $\lambda_{ij}$ as free parameters, including the
nondiagonal ones, since it is reasonable to expect that
$|\lambda_{ii}|\leq 1$ in a general context.

\begin{figure}
\centering
\includegraphics[width=3.5in]{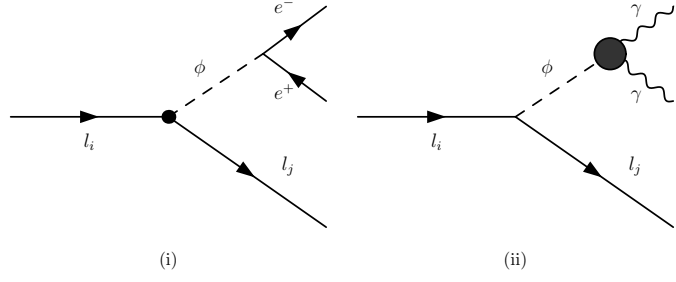}
\caption{\label{FIG1}Feynman diagrams contributing to the $l_i\to
l_je^+e^-$ and $l_i\to l_j \gamma \gamma$ decays. The large dot
represents a fermion loop.}
\end{figure}

\begin{figure}
\centering
\includegraphics[width=3.5in]{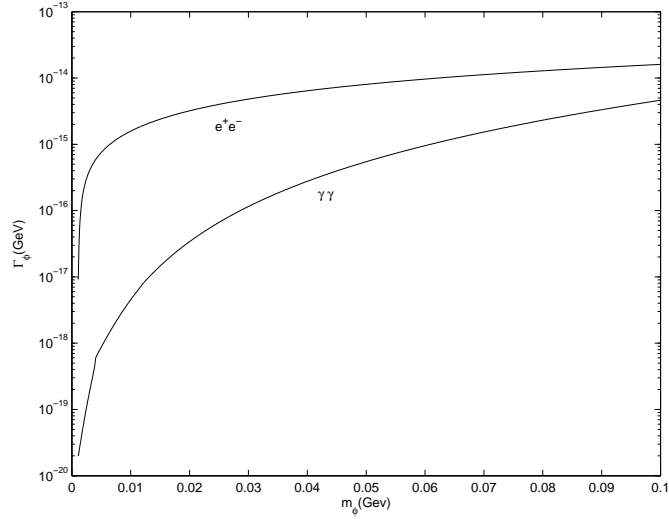}
\caption{\label{FIG2}Decay widths for the modes $\phi\to e^+e^-$ and
$\phi\to \gamma \gamma$ as function of $m_\phi$.}
\end{figure}

We now proceed to analyze the $l_i\to l_j\gamma \gamma$ and $l_i\to
l_j e^+e^-$ decays. From the Feynman diagrams depicted in Fig.
\ref{FIG1}, the branching fractions can be written, in the narrow
width approximation, as
\begin{equation}
B(l_i\to l_j\gamma \gamma)=B(l_i\to l_j\phi)B(\phi \to \gamma
\gamma),
\end{equation}
\begin{equation}
B(l_i\to l_je^+e^-)=B(l_i\to l_j\phi)B(\phi \to e^+e^-).
\end{equation}
Note that apart from the Feynman diagram shown in Fig.
\ref{FIG1}(ii), the decay $l_i\to l_j\gamma \gamma$  can also
proceed via those reducible graphs in which one of the photons is
emitted from an external lepton, or also via box diagrams. However,
general considerations suggest that this class of contributions is
marginal \cite{SHALOM}. From the above expressions, it follows that
\begin{equation}
\label{keyeq} B(l_i\to l_j\gamma \gamma)=\frac{\Gamma(\phi \to
\gamma \gamma)}{\Gamma(\phi \to e^+e^-)}B(l_i\to l_j e^+e^-),
\end{equation}
where
\begin{equation}
\Gamma(\phi \to
e^+e^-)=\frac{\alpha|\lambda_{ee}|^2m_\phi}{2s^2_{2W}}\left(\frac{m_e}{m_Z}\right)^2
\left(1-\frac{4m^2_e}{m^2_\phi}\right)^{3/2},
\end{equation}
with $s_{2W}=2\sin\theta_W\cos\theta_W$, being $\theta_W$ the weak
angle. On the other hand, the decay width for the two photon mode is
given by \cite{Resnick:1973vg}
\begin{equation} \Gamma(\phi \to \gamma
\gamma)=\frac{\alpha^3m_\phi}{16\pi^2s^2_{2W}}\left(\frac{m_\phi}{m_Z}\right)^2|F|^2,
\end{equation}
with
\begin{equation}
F=\sum_{f=l,q} N^f_CQ^2_f\lambda_{ff}xf(x),
\end{equation}
and
\begin{equation}
f(x)=\left\{\begin{array}{ll}
\left(\arcsin\frac{1}{\sqrt{x}}\right)^2 &\quad x\geq 1, \\
-\left(\arccos{\rm
h}\frac{1}{\sqrt{x}}-\frac{i\pi}{2}\right)^2&\quad
x<1,\\
\end{array} \right.
\end{equation}
where $x=4m^2_f/m^2_\phi$, $N^f_C$ is the color index, and $Q_f$ is
the electric charge in units of the positron charge. For the sake of
illustration, we have evaluated the decay widths $\Gamma(\phi \to
e^+e^-)$ and $\Gamma (\phi\to \gamma \gamma)$ for $\lambda_{ff}=1$.
They are shown as function of $m_\phi$ in Fig. \ref{FIG2}. We can
observe that $\Gamma_\phi$ is of order $10^{-14}$ GeV at most, which
means that the narrow width approximation used in obtaining Eq.
(\ref{keyeq}) is a very good approximation indeed because
$\Gamma_\phi$ is thirteen orders of magnitude smaller than $m_\phi$.
Also, we can see that, depending on the value of $m_\phi$, $\Gamma
(\phi\to \gamma \gamma)$ is up to three or one order of magnitude
smaller than $\Gamma(\phi \to e^+e^-)$. Although in some particular
models $\Gamma (\phi\to \gamma \gamma)$ may reach values near
$\Gamma(\phi \to e^+e^-)$ \cite{TAVARES}, it is reasonable to assume
that  $\Gamma (\phi\to \gamma \gamma)<\Gamma(\phi \to e^+e^-)$ as
the $\gamma \gamma$ mode is always a loop-generated effect in a
renormalizable theory. This means that Eq. (\ref{keyeq}) can be
written as an inequality
\begin{equation}
\label{keyeq2}
 B(l_i\to l_j\gamma \gamma)<B(l_i\to l_j e^+e^-),
\end{equation}
which is in agreement with the fact that the decay $l_i\to l_j\gamma
\gamma$  can only arise at one-loop or higher orders, whereas the
$l_i\to l_j e^+e^-$ reaction can be induced at the tree level. It is
worth emphasizing that (\ref{keyeq2}) is only true under the
assumption that $\lambda_{ee}\simeq 1$. If the very light $\phi$ was
leptophobic, it would only decay into a photon pair, in which case
the inequality (\ref{keyeq2}) would not hold.

The current experimental constraints on $B(l_i\to l_j e^+e^-)$
\cite{PDG} allow us to translate (\ref{keyeq2}) into the following
bounds on the  decays $l_i\to l_j\gamma \gamma$:
\begin{eqnarray}
B(\mu^-\to e^-\gamma \gamma)&<&1.0\times 10^{-12}, \\
B(\tau^-\to e^-\gamma \gamma)&<&2.0\times 10^{-7}, \\
B(\tau^-\to \mu^-\gamma \gamma)&<&1.9\times 10^{-7}.
\end{eqnarray}
Notice that the constraint on $B(\mu^-\to e^-\gamma \gamma)$ is
almost two orders of magnitude more stringent than the experimental
one \cite{PDG}: $B_{\rm Exp.}(\mu^-\to e^-\gamma \gamma)<7.2\times
10^{-11}$.

\begin{figure}
\centering
\includegraphics[width=3.5in]{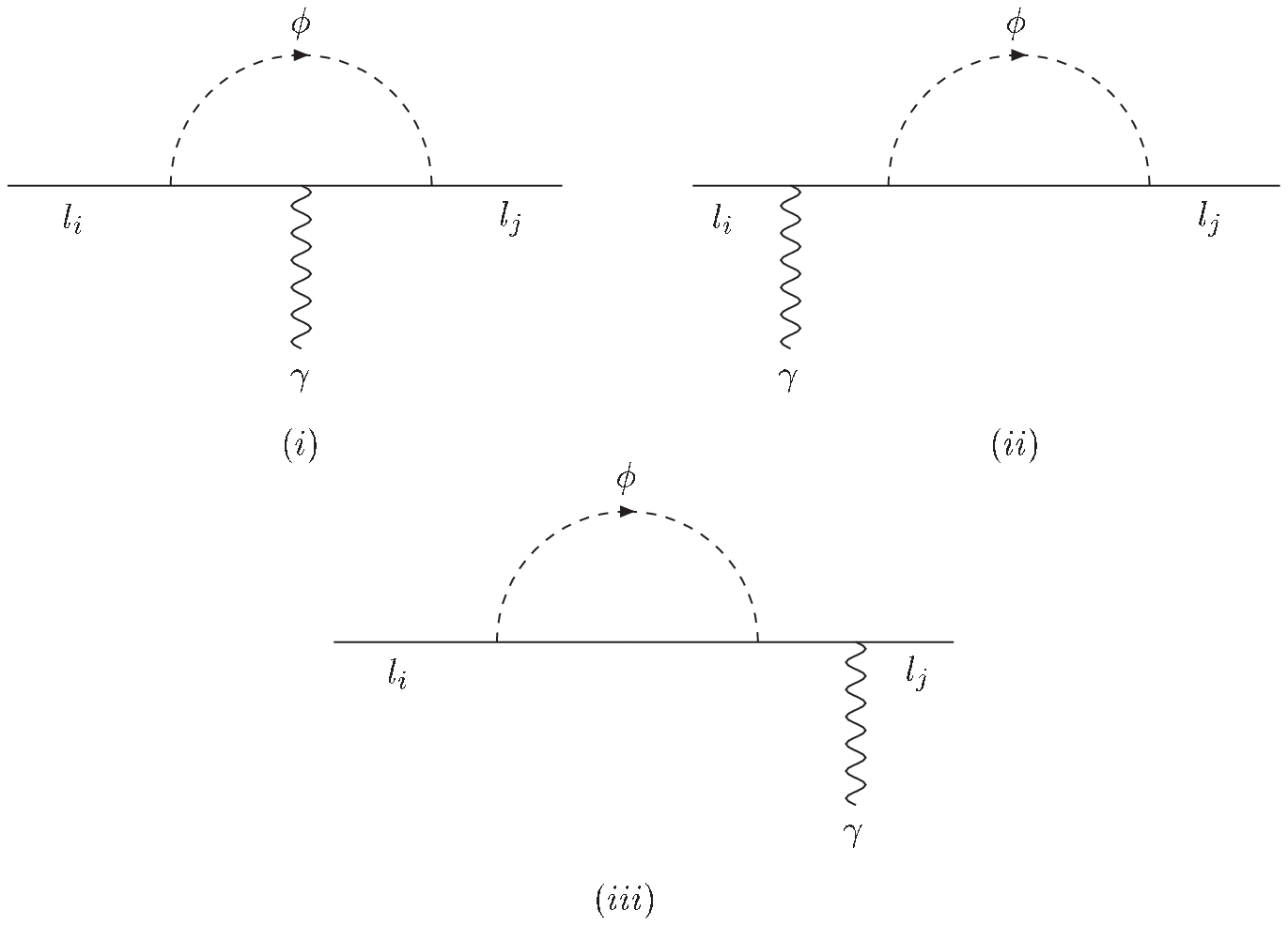}
\caption{\label{FIG3}Feynman diagrams contributing to the $l_i\to
l_j\gamma$ decay.}
\end{figure}

The size of the parameters $\lambda_{ij}$  can be estimated using
the current experimental constraints on the decay $l_i\to
l_j\gamma$. It proceeds via the loop diagrams shown in Fig.
\ref{FIG3}, which involve the pseudoscalar boson and the charged
leptons. The branching ratio can be written as
\begin{equation}
\label{li->ljg} B(l_i\to
l_j\gamma)=\frac{\alpha^3}{16\pi^2s^4_{2W}}\left(\frac{m_j}{\Gamma_i}\right)
\left(\frac{m_i}{m_Z}\right)^2|\sum_k\left(\frac{m_k}{m_Z}\right)\lambda_{ik}\lambda_{kj}|^2|{\cal
A}_k|^2,
\end{equation}
where $\Gamma_i$ is the $l_i$ total width and
\begin{equation}
{\cal
A}_k=-\frac{1}{2}+\left(\frac{m_i}{m_k}-1\right)m^2_kC_0(m^2_i,0,0,m^2_k,m^2_\phi,m^2_k)
+\frac{m_k(m_k-m_i)-m^2_\phi}{m^2_i}\Big[B_0(0,m^2_k,m^2_\phi)-B_0(m^2_i,m^2_k,m^2_\phi)\Big].
\end{equation}
with the Passarino-Veltman scalar functions given by

\begin{equation}
B_0(0,m^2_k,m^2_\phi)-B_0(m^2_i,m^2_k,m^2_\phi)=1+
\frac{\xi}{m_i^2}\,{\rm arccosh}\left(\frac{m_k^2 - m_i^2 +
m_{\phi}^2}{2\,m_k\,m_{\phi}} \right) - \frac{1}{2}\left(\frac{m_k^2
- m_{\phi}^2}{m_i^2} - \frac{ m_k^2 + m_{\phi}^2}{m_k^2 -
m_{\phi}^2} \right) \log\left(\frac{m_k^2}{m_{\phi}^2}\right),
\end{equation}

\begin{equation}
C_0(m^2_i,0,0,m^2_k,m^2_\phi,m^2_k)=\frac{1}{m_i^2}\left[{\rm
Li}_2\left(1-\frac{m_{\phi}^2}{m_k^2}\right) -{\rm
Li}_2\left(\frac{2\,m_i}{\eta+\xi}\right)-{\rm
Li}_2\left(\frac{2\,m_i}{\eta-\xi}\right)\right],
\end{equation}
with $\xi^2=\left(m_k^2+m_i^2-m_{\phi}^2\right)^2- 4m_k^2m_i^2$ and
$\eta=m_k^2+m_i^2-m_{\phi}^2$.

The analysis of the loop amplitude shows that the dominant
contribution comes from $l_k=l_i$. In such a case, one can take
${\cal A}_k\approx -1/2$. Eq. (\ref{li->ljg}) thus  translates into
the following bound when the experimental constraint is taken into
account
\begin{equation}
|\lambda_{ii}|^2|\lambda_{ij}|^2<\frac{64\pi^2s^4_{2W}}{\alpha^3}\left(\frac{\Gamma_i}{m_i}\right)\left(\frac{m_Z}{m_i}\right)^3
\left(\frac{m_Z}{m_j}\right)B_{Exp.}\left(l_i\to l_j\gamma\right).
\end{equation}
Using the experimental constraints on the $l_i\to l_j\gamma$ decays
\cite{PDG} we obtain
\begin{eqnarray}
|\lambda_{\mu e}||\lambda_{\mu \mu}|&<&1.79\times 10^{-3}, \\
|\lambda_{\tau \mu}||\lambda_{\tau \tau}|&<&1.95\times 10^{-1}, \\
|\lambda_{\tau e}||\lambda_{\tau \tau}|&<&3.15
\end{eqnarray}

It is worth examining the bounds that can be obtained from the
three-body decays $l_i\to \l_j\, e^-e^+$ and $l_i\to \gamma \gamma$.
The branching ratio of the former decay can be written as
\begin{equation}
B(l_i\to \l_j\, e^-e^+)=\frac{\alpha^2}{8 \pi
{s_{2W}^4}}\left(\frac{m_j}{\Gamma_i}\right)\left(\frac{m_i}{m_Z}\right)^2
\left(\frac{m_e}{m_Z}\right)^2|\lambda_{ij}|^2|\lambda_{ee}|^2\,\int^1_0
\frac{x(1-x)^2}{(x-y)^2+ yz} dx,
\end{equation}
with $x=p^2/m_i^2$, $y=m_\phi^2/m_i^2$, and $z=\Gamma_\phi^2/m_i^2$,
$p$ being the four-momentum of the exchanged pseudoscalar. For
$m_\phi\simeq m_\mu/2$, the experimental bounds \cite{PDG} yield
\begin{eqnarray}
|\lambda_{\mu e}||\lambda_{ee}|&<& 3\times 10^{-6},\\
|\lambda_{\tau \mu}||\lambda_{ee}|&<& 4.5,\\
|\lambda_{\tau e}||\lambda_{ee}|&<& 66.0.
\end{eqnarray}
These upper bounds depend smoothly on the $\phi$ mass and do not
deviate significantly from the above values in the interval $0.01$
GeV $\le m_\phi\le 0.125$ GeV, except for the bound on
$|\lambda_{\mu e}||\lambda_{ee}|$, which relaxes up to $10^{-5}$
when $m_\phi\simeq 0.01$ GeV.

As for the three-body decay $l_i \to l_j\gamma\gamma$, Ref.
\cite{PDG} only reports an experimental bound on $\mu\to e
\gamma\gamma$. Assuming the existence of a very light pseudoscalar,
the corresponding branching ratio can be computed  via the narrow
width approximation: $B(\mu\to e\gamma\gamma)\simeq B(\mu\to e
\phi)B(\phi\to\gamma\gamma)$, where

\begin{equation}
B(\mu\to e\phi)\simeq\frac{\alpha\,|\lambda_{\mu e}|^2}{4
s^2_{2W}}\left(\frac{m_e}{\Gamma_\mu}\right)
\left(\frac{m_\mu}{m_Z}\right)^2\left(1-y_\mu^2\right)^2,
\end{equation}
with $y_\mu=m_\phi/m_\mu$. Numerical calculation, when combined with
the experimental constraint on $B(\mu\to e \gamma\gamma)$, yields

\begin{equation}
|\lambda_{\mu e}|<2.14\times 10^{-8},
\end{equation}
for $m_\phi=m_\mu/2$. This bound is stronger than the one obtained
from the two-body decay $\mu\to e \gamma$ as $\lambda_{ee}$ is
expected to be of order unity at most. This stems from the fact that
the three-body decay $\mu \to e \gamma\gamma$ becomes significantly
enhanced on resonance of the $\phi$ boson.

We now would like to comment on some realistic  models suited for
the class of effects that we just have analyzed. Specific examples
of models which allow a very light pseudoscalar are THDMs. It was
shown recently \cite{TAVARES} that although the parameter space of
this class of models has been tightly constrained by the most recent
measurements on the muon anomalous magnetic moment and other low
energy processes, the existence of a very light scalar with a mass
at the MeV level can still be possible if some parameters of the
model are fine-tuned. As has been pointed out in Ref.
\cite{Dobrescu}, other examples of theories with extended Higgs
sectors that allow a very light pseudoscalar are the minimal
composite Higgs model (MCHM) \cite{MCHM} and the next-to minimal
supersymmetric standard model (NMSSM) \cite{NMSSM}. In both of these
models a light pseudoscalar is allowed because its mass is
controlled by the explicit breaking of a spontaneously broken $U(1)$
symmetry. The pseudoscalar is an axion due to the fact that the
global $U(1)$ symmetry has a QCD anomaly. Current bounds on the mass
of such a pseudoscalar are at the MeV level.

Lepton flavor violating transitions can be considerably suppressed
if they are mediated by a very heavy particle and the involved
couplings are negligible, but they can be favored if either of these
conditions are sufficiently relaxed. For instance, in the Cheng-Sher
ansazt is assumed that these effects can be mediated by a relatively
light scalar, with a mass of the order of the Fermi scale, at the
same time that weak LFV couplings are postulated. In this work, the
Cheng-Sher ansazt spirit was retained, but the requirement of a
relatively light scalar boson was maximally relaxed, admitting the
possible existence of a very light pseudoscalar boson. An inequality
between the branching ratios for the three-body decays $l_i\to
l_j\gamma \gamma$ and $l_i\to l_j e^+e^-$ [Eq. (\ref{keyeq})] arises
as a new ingredient introduced by the long-lived nature of this
particle. As far as the strength of the $\phi l_il_j$ vertex is
concerned, current experimental data show that, for leptons of
adjacent families ($\phi \mu e$ and $\phi \tau \mu$), it is still
more suppressed than expected for a Higgs boson with mass of the
order of the Fermi scale.

\acknowledgments{We acknowledge support from Conacyt under grant
U44515-F.}


\begin{thebibliography}{99}
\bibitem{HBLEP} A. Heister {\it et al.} Phys.
Lett. B {\bf 526}, 191 (2002).

\bibitem{AXIONS} S. Weinberg, Phys. Rev. Lett. {\bf 40}, 223
(1978); F. Wilczek, Phys. Rev. Lett. {\bf 40}, 279 (1978).

\bibitem{PQ}R. D. Peccei and H.R. Quinn, Phys. Rev. Lett. {\bf 38},
1440 (1977); Phys. Rev. D {\bf 16}, 1791 (1977).

\bibitem{FAM} F. Wilczek, Phys. Rev. Lett. {\bf 49}, 1549 (1982).

\bibitem{MAJ} Y. Chikashige, R. N. Mohapatra, and R. D. Peccei, Phys.
Lett. B {\bf 98}, 265 (1981); G. B. Gelmini and M. Roncadelli, Phys.
Lett. B {\bf 99}, 411 (1981).

\bibitem{SHER} D. L. Anderson, C. D. Carone, and M. Sher, Phys. Rev.
D {\bf 67}, 115013 (2003).

\bibitem{TAVARES} F. Larios, G. Tavares-Velasco, and C. P. Yuan,
Phys. Rev. {\bf D64}, 055004 (2001); Phys. Rev. D {\bf 66}, 075006
(2002).

\bibitem{SHER1} T. P. Cheng and M. Sher, Phys. Rev. D {\bf 35},
3484.

\bibitem{SHER2} M. Sher and Y. Yuan, Phys. Rev. D {\bf 44}, 1461 (1991).

\bibitem{FUKUDA} Y. Fukuda {\it et al.}
Phy. Rev. Lett. {\bf 77}, 1683 (1996).

\bibitem{RECENTWORKS} S. Nie and M. Sher, Phys. Rev. D {\bf 58},
097701 (1998); J. L. D\'{\i}az-Cruz and J. J. Toscano, Phys. Rev.
{\bf D62}, 116005 (2000); M. Sher, Phys. Lett. B {\bf 487}, 151
(2000); T. Han and D. Marfatia, Phys. Rev. Lett. {\bf 86}, 1442
(2001); D. Black, T. Han, H-J. He, and M. Sher, Phys. Rev. D {\bf
66}, 053002 (2002);  A. Atre, V. Barger, and T. Han, Phys. Rev. D
{\bf 71}, 113014 (2005); E. Arganda, Ana M. Curiel, M. J. Herrero,
and D. Temes, Phys. Rev. D {\bf 71}, 035011 (2005); S. Kanemura, T.
Ota, and K. Tsumura, arXiv: {hep-ph/0505191}; J. L. D\'{\i}az-Cruz,
R. Noriega-Papaqui, and A. Rosado, Phys. Rev. {\bf D71}, 015014
(2005).


\bibitem{SHALOM} A. Gemintern, S. Bar-Shalom, G. Eilam, and F.
Krauss, Phys. Rev. D {\bf 67}, 115012 (2003).

\bibitem{Resnick:1973vg}
  L.~Resnick, M.~K.~Sundaresan, and P.~J.~S.~Watson,
  Phys.\ Rev.\ D {\bf 8}, 172 (1973).


\bibitem{PDG} S. Eidelman {\it et al.}, Phys. Lett. B {\bf 592}, 1
(2004).

\bibitem{Dobrescu}B. Dobrescu, G. Landsberg, and K. Matchev, Phys. Rev. D 63,
075003 (2001).

\bibitem{MCHM}B. A. Dobrescu, Phys. Rev. D 63, 015004 (2001).

\bibitem{NMSSM}B. A. Dobrescu and K. T. Matchev, J. High Energy Phys. 09,
031 (2000).

\end{thebibliography}
\end{document}